\documentclass[prl,reprint,amsmath,amssymb,preprintnumbers,superscriptaddress,nofootinbib]{revtex4-1}

\usepackage{epsfig,epstopdf}
\usepackage{subfigure}
\usepackage{color}

\def\beq{\begin{equation}}
\def\eeq{\end{equation}}
\def\bea{\begin{eqnarray}}
\def\eea{\end{eqnarray}}

\newcommand{\lsim}{
\mathrel{\hbox{\rlap{\hbox{\lower4pt\hbox{$\sim$}}}\hbox{$<$}}}}
\newcommand{\gsim}{
\mathrel{\hbox{\rlap{\hbox{\lower4pt\hbox{$\sim$}}}\hbox{$>$}}}}

\newcommand{\dis}[1]{\begin{equation}\begin{split}#1\end{split}\end{equation}}
\newcommand{\mpl}{m_{\rm Pl}}

\begin{document}

\title{The entropic quasi-de Sitter instability time from the distance conjecture}

\author{Min-Seok Seo}
\email{minseokseo57@gmail.com}
\affiliation{Department of Physics Education, Korea National University of Education,
\\ 
Cheongju 28173, Republic of Korea}

\begin{abstract}
\noindent 
From the entropy argument for the dS swampland conjecture which connects the Gibbons-Hawking entropy bound with the distance conjecture, we find the entropic quasi-dS instability time given by $1/(\sqrt{\epsilon_H}H)\log(\mpl/H)$ as the lifetime of quasi-dS spacetime. 
It depends on the slow-roll parameter, and  contains the logarithmic factor $\log(\mpl/H)$ which can be found in the scrambling (or decoherence) time as well.
Such a logarithmic factor enhances the geodesic distance of the modulus from the mere Planck scale, and also possibly relaxes the bound on $\mpl^2 \nabla^2 V/V$.

\end{abstract}
\maketitle

\section{Introduction and Summary}
 
  Over the last decade, various quantum gravity constraints on the low energy effective field theory (EFT) have been conjectured under the name of `swampland program' (for reviews, see \cite{Brennan:2017rbf, Palti:2019pca}).
  Whereas these conjectures are motivated by string theory model building, it was soon realized that some of them are supported by more generic quantum gravity argument, and even connected with each other.
  In the recent de Sitter (dS) swampland conjecture \cite{Obied:2018sgi}, for example, the instability of dS spacetime was argued \cite{Ooguri:2018wrx} based on Bousso's covariant entropy bound \cite{Bousso:1999xy} and the distance conjecture \cite{Ooguri:2006in}.
 More concretely, the distance conjecture tells us that as a modulus traverses along the trans-Planckian geodesic distance, towers of states descend from UV, resulting in the rapid increase in the entropy.
  However, the entropy cannot exceed the Gibboons-Hawking entropy bound  $S_{\rm GH}=\pi \mpl^2 /H^2$ \cite{Gibbons:1977mu} so once the entropy saturates $S_{\rm GH}$ we expect that it soon decreases.
  This resembles the negative temperature system, in which the entropy density is maximized at some finite energy density \cite{Seo:2019mfk}.
  Such a feature of the entropy is realized by taking an ansatz $S=N^p (H/\mpl)^q$ with $p>0$ and $q+2>0$ \cite{comment0}.
  From this, we impose the condition $S \lesssim S_{\rm GH}$, or
  \dis{N^p \Big(\frac{H}{\mpl}\Big)^{2+q} \lesssim 1,\label{Eq:ineq}}
  where both $N$ and $H$ depend on the modulus $\phi$.
  By varying this with respect to $\phi$ we obtain the bound at the saturation \cite{Ooguri:2018wrx}  (see also \cite{Seo:2018abc}),
  \dis{-\frac{1}{H}\frac{d H}{d\phi} \gtrsim \frac{p}{q+2}\frac{1}{N}\frac{dN}{d\phi},}
  which can be rewritten as the bound on  $\epsilon_H=-\dot{H}/H^2=\dot{\phi}^2/(2\mpl^2 H^2)$,
  \dis{\sqrt{\epsilon_H}\gtrsim \frac{\sqrt2 p}{q+2}\frac{\mpl}{N}\frac{dN}{d\phi} \equiv \sqrt{\epsilon_c}.\label{Eq:dSbound}}
 Hence the increase rate in $N$, $(\mpl/N) (dN/d\phi)$ of order one as postulated by the distance conjecture is compensated by the order one decrease rate in $H$, $\epsilon_H\sim {\cal O}(1)$.

    The entropy argument above shows that quasi-dS spacetime satisfying $\epsilon_H \ll 1$ is allowed by quantum gravity only for the sub-Planckian geodesic distance $\Delta \phi$ over which UV degrees of freedom are heavy enough to decouple from EFT.
    If the energy density is dominated by the concave downward potential such that $\epsilon_H$ gradually increases from zero as the potential height decreases, $\phi$ slow-rolls down from the top of the potential ($\epsilon_H=0$) until $\epsilon_H$ saturates the bound $\epsilon_c$.
  Then the period of slow-roll can be interpreted as a lifetime of quasi-dS spacetime, which we will call the `entropic quasi-dS instability time',  $t_{\rm edS}$.
 When we identify $\phi$ with the inflaton in the early Universe, $t_{\rm edS}$ corresponds to the period of the inflation.

Indeed, we can find various characteristic time scales concerning perfect dS spacetime ($\epsilon_H=0$),  just like the Compton wavelength or the classical radius of electron in quantum electrodynamics. 
 The most evident one might be the horizon size, $H^{-1}$.
 It is also interpreted as the `classical break-time' after which the linearity is no longer a good description for the metric fluctuation, graviton \cite{Dvali:2017eba}.
 On the other hand, the non-linear self-interaction between gravitons eventually deforms the (minimal uncertainty) coherent state, which becomes apparent after the `quantum break-time' $\mpl^2/H^3$ \cite{Dvali:2017eba}.
 In addition, motivated by the black hole thermodynamics, we find the `scrambling time', the time scale for the internal quantum state to evolve into the mixed state \cite{Hayden:2007cs, Sekino:2008he}, given by $(1/H)\log(\mpl/H)$.
 If dS sacetime lives shorter than the scrambling time as conjectured in \cite{Bedroya:2019snp}, trans-Planckian mode cannot escape the horizon, and   the past lightcone of dS states within the stretched horizon cannot decouple from the initial hypersurface  \cite{Nomura:2011dt}.
  Since it takes the scrambling time for the super-horizon graviton modes to lose their quantum nature through the interaction with the sub-horizon modes  \cite{Gong:2019yyz} (see also \cite{Zurek:1994wd}), the scrambling time is also called a decoherence time scale.
 
  For quasi-dS spacetime, there are more time scales which depend on $\epsilon_H$.
  The non-zero $\epsilon_H$ parametrizes the spontaneous breaking of dS isometry, from which the trace part of the metric fluctuation becomes physical  by absorbing the  inflaton fluctuation \cite{Cheung:2007st} (see also \cite{Prokopec:2010be, Gong:2016qpq}).
 The decoherence time scale for such a `curvature perturbation' is given by $(1/H)\log[\mpl/(\sqrt{\epsilon_H}H)]$ \cite{Nelson:2016kjm}, in which the $\epsilon_H$ dependence comes from the coupling between  sub- and super-horizon modes given by $\epsilon_H (H/\mpl)^2$.
 The entropic quasi-dS instability time  presented in this work provides another logarithmic time scale, reflecting variation of $\epsilon_H$ during the  slow-roll of the inflaton.

\section{The entropic quasi-dS instability time} 

 To obtain the entropic quasi-dS instability time, we consider the inflaton slowly rolling down the concave downward potential, in which $\epsilon_H$ at initial time $t_i$ is negligibly small.
   Let the bound \eqref{Eq:ineq} is saturated at $t_f$, such that $\epsilon_H(t_f)=\epsilon_c$ given by \eqref{Eq:dSbound}.
 we also take an ansatz for  $N$ as  \cite{Ooguri:2018wrx},
 \dis{N=N_0(\phi) e^{\gamma \Delta \phi/\mpl}.}
 We assume here that $N_0$ is almost constant, $(\mpl/N_0)(dN_0/d\phi) \ll \gamma$, such that $(\mpl/N)(dN/d\phi)\simeq \gamma$ which is ${\cal O}(1)$ as the distance conjecture predicts.
 From   $\epsilon_H=\dot{\phi}^2/(2\mpl^2 H^2)$ we write   $\Delta \phi$ as
 \dis{\Delta \phi =\int_{t_i}^{t_f}\dot{\phi} dt=\int_{t_i}^{t_f}\sqrt{2\epsilon_H} \mpl H dt.}
  To make the physical meaning more clear, we define the `time average' over the entropic quasi-dS instability time $t_{\rm edS} =t_f-t_i$ of any time-dependent quantity $X(t)$ by
  \dis{\langle X \rangle \equiv \frac{1}{t_{\rm edS}}\int_{t_i}^{t_f}X(t) dt.}
  Then we have the simple expression,
  \dis{\Delta \phi=\sqrt2 \mpl \langle\sqrt{\epsilon_H}H\rangle t_{\rm edS} ,\label{Eq:geodist}}
  from which we obtain
  \dis{t_{\rm edS} &\simeq
  \frac{1}{\sqrt2 \gamma p  \langle\sqrt{\epsilon_H}H\rangle}\Big[(q+2)\log\Big(\frac{\mpl}{H_f}\Big)-p\log N_0\Big]
  \\
  &\simeq \frac{(q+2)}{\sqrt2 \gamma p  \langle\sqrt{\epsilon_H}H\rangle}\log\Big(\frac{\mpl}{H_f}\Big).}
 Here $H_f \equiv H(t_f)$, the value of the Hubble parameter at the saturation of inequality \eqref{Eq:ineq} and $H_f\ll \mpl$ is assumed for the last equality.
 Putting this into \eqref{Eq:geodist}, we obtain
 \dis{\Delta \phi\simeq \mpl\frac{q+2}{\gamma p}\log\Big(\frac{\mpl}{H_f}\Big).}
 That is,  $\Delta \phi$ is not simply of order of $\mpl$ but enhanced by the factor $\log(\mpl/H)$.
 Such an enhancement of course comes from equating $N$  in which $\Delta \phi$ is exponentiated with the some positive power of $\mpl/H$.
 This is a typical prediction of the distance conjecture \cite{Scalisi:2018eaz} (see also \cite{Palti:2019pca}).

 \begin{figure}[!h]
  \begin{center}
   \includegraphics[width=0.40\textwidth]{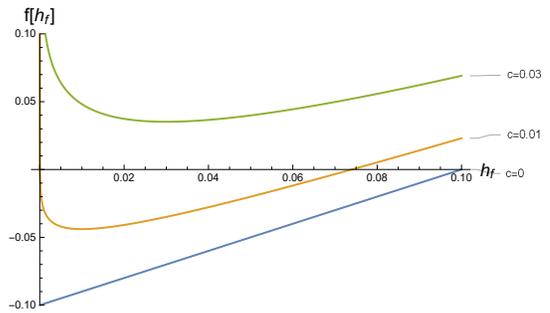}
  \end{center}
 \caption{The plot for $f(h_f)\equiv h_f-h_i-c\log h_f$ with $h_i$ is held fixed at $0.1$.
  }
\label{Fig:hi01}
\end{figure}
 \begin{figure}[!h]
  \begin{center}
   \includegraphics[width=0.45\textwidth]{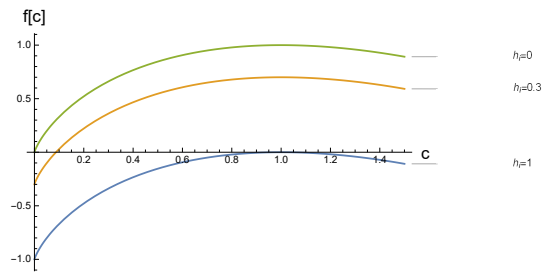}
  \end{center}
 \caption{The plot for ${\rm min}[f(h_f)]=f(c)$ with $h_i=0$, $0.3$, and $1$ (from above).
 All these curves are maximized at $c=1$.
  }
\label{Fig:fcvsc}
\end{figure}

 If $\phi$ spends most of time on the region of the potential  satisfying $\epsilon_H \ll 1$,  $H$ is almost constant so we can approximate $\langle\sqrt{\epsilon_H}H\rangle$ by $\langle\sqrt{\epsilon_H}\rangle H_f$, which is much smaller than $H_f$.
 In this case, $t_{\rm edS}$ is much longer than the scrambling time $(1/H)\log(\mpl/H)$.
 These two time scales are comparable provided $\epsilon_H$ rapidly departs from zero, but then $H$ cannot be approximated as a constant any longer.
 We also note that for the $\epsilon_H \ll 1$ region dominant case, the number of $e$-folds is given by
 \dis{N_e\simeq H_f t_{\rm edS}\simeq \frac{(q+2)}{\sqrt2 \gamma p}\frac{1}{\sqrt{\epsilon_H}}\log\Big(\frac{\mpl}{H_f}\Big),}
 which is typically large enough to accommodate almost possible inflation models.
 This can be shown from the fact that the ratio $\mpl/H$ is measured by the power spectrum of the tensor perturbation $\Delta_h^2=(2/\pi^2)(H/\mpl)^2$, which is connected to  $\epsilon_H$ through $H/\mpl=(\pi/\sqrt2)\sqrt{r}\Delta_{\cal R}$ where   $r=\Delta_h^2/\Delta_{\cal R}^2=16\epsilon_H$ is the tensor-to-scalar ratio.
 Since the power spectrum of the curvature perturbation has the size $\Delta_{\cal R}\sim 10^{-5}$, for $r\sim 0.1$ (or equivalently $\epsilon_H\sim 0.01$), the number of $e$-folds is estimated to be $N_e \sim 10^2$. 
 
 Of course, for sizeable $\epsilon_H$, $\langle\sqrt{\epsilon_H}H\rangle$ is no longer approximated by $\langle\sqrt{\epsilon_H}\rangle H_f$ and its value depends on the shape of the potential as well as the value of $H_i$.  
 To see how $H_i$ determines $H_f$, we investigate the relation between $H_i \equiv H(t_i)$, $H_f$ and $\epsilon_H(\phi)$ using the definition of $\epsilon_H$,
 \dis{\frac{H_i-H_f}{\mpl}&=\frac{\langle\epsilon_H H^2\rangle t_{\rm edS}}{\mpl}
 \\
 &=\frac{q+2}{\sqrt2 \gamma p}\frac{\langle\epsilon_H H^2\rangle}{\langle\sqrt{\epsilon_H} H \rangle \mpl}\log\Big(\frac{\mpl}{H_f}\Big).}
 Rewriting this in terms of dimensionless quantities $h_f \equiv H_f/\mpl$, $h_i \equiv H_i/\mpl$, and $c\equiv [(q+2)/(\sqrt2 \gamma p)][\langle\epsilon_H H^2\rangle/(\langle\sqrt{\epsilon_H} H \rangle \mpl)] \sim \sqrt{\epsilon_H}(H/\mpl)$ all of which are smaller than one, we find that the solution $h_f$ exists if the graph for $f(h_f)\equiv h_f-h_i-c\log h_f$ which is bounded from below (see Fig. \ref{Fig:hi01}) crosses zero.
 That means, the minimum of $f(h_f)$, ${\rm min}[f(h_f)]= f(c)$ must be negative.
 Whereas  in this case there are two possible solutions for $h_f$, only the biggest one  is physical since it has a smooth limit $h_f \to h_i$ as $c\to 0$ ($H$ does not change for the flat potential).
 As $c$ gets smaller, $h_f$ approaches to $h_i$.
 We also find from Fig. \ref{Fig:fcvsc} that the smaller the value of $h_i$, the smaller values of $c$ are allowed and $h_f$ becomes close to $h_i$. 

  Another consequence of the logarithmic enhancement in $\Delta \phi$ is the relaxation of the bound on $\mpl^2 \nabla^2 V/V$.
  In the refined dS swampland conjecture \cite{Andriot:2018wzk, Garg:2018reu, Ooguri:2018wrx}, it was pointed out that the mere order one bound on $\mpl \nabla V/V$, or equivalently $\epsilon_H \sim {\cal O}(1)$ in the case of slow-roll approximation is not sufficient to describe the dS instability.
  As we have seen, the entropy argument implies that quasi-dS spacetime can exist until the entropy saturates $S_{\rm GH}$ \cite{Ooguri:2018wrx}.
  For dS spacetime to be unstable, we need an additional condition that even though the existence of the point $\phi_0$ at which the potential is flat ($\nabla V(\phi_0)=0$) is not excluded by quantum gravity, spacetime immediately deviates from dS  by having $\mpl^2 \nabla^2 V/V \sim -{\cal O}(1)$ at $\phi_0$.
  In the case of the concave downward potential, the absolute value $\mpl^2 |V''|/V$  at the top of the potential (at $\phi_0$) can be estimated  by considering the change in $V'$ over the range $[\phi_0, \phi_0+\Delta \phi]$ :  using $\epsilon_H\simeq \epsilon_V=(\mpl^2/2)(V'/V)^2$ in the slow-roll approximation, from $V'(\phi_0)=0$ and $\epsilon_H(\phi_0+\Delta \phi)=\epsilon_c$ which is given by \eqref{Eq:dSbound}, we obtain
 \dis{\mpl^2 \frac{|\Delta V'|}{\langle V\rangle}&=\mpl^2 \frac{V'(\phi_0+\Delta \phi)}{\langle V\rangle \Delta \phi}\simeq \mpl\frac{\sqrt{2\epsilon_c} V(\phi_0+\Delta \phi)}{\langle V\rangle \Delta \phi}
 \\
 &\simeq 2\Big(\frac{\gamma p}{q+2}\Big)^2\frac{H_f^2}{\langle H^2\rangle\log(\mpl/H_f)},\label{Eq:V''}}
 where $V=3\mpl^2 H^2$ is used.
 Hence the logarithmically enhanced $\Delta \phi$ indicates that the slope increases from zero to $\epsilon_c$  more moderately allowing the smaller $\mpl^2 |V''|/V$.
 Especially, for the concave downward potential, $\mpl^2 |V''|/V$ is maximized at $\phi_0$ so \eqref{Eq:V''} can be the lower bound on $\mpl^2 (|V''|/V)(\phi_0)$,  which can be smaller by   the enhanced $\Delta \phi$.
 However, we do not have a concrete example of a potential for which the relaxed bound on $m_{\rm Pl}^2|V''|/V$ makes a difference: typical Higgs or axion potentials already give $m_{\rm Pl}^2|V''|/V\sim -O(1)$.

\section{ Discussion}

In the presence of the horizon with the radius $1/H$, the quantum fluctuation becomes classical after its wavelength stretches beyond the horizon.
 If the period of the inflation is long enough, even the trans-Planckian modes escape the horizon at the late stage of the inflation, then we may need an unknown new physics for the complete description \cite{Martin:2000xs}.
 Recently proposed trans-Planckian censorship conjecture \cite{Bedroya:2019snp} avoids such a difficulty by requiring that dS lifetime is shorter than the scrambling time, $(1/H)\log(\mpl/H)$ \cite{comment}. 
 Regarding the entropic quasi-dS instability time, since $\epsilon_H <{\cal O}(1)$ until just before $t_{\rm edS}$, we have $\langle \sqrt{\epsilon_H} H\rangle < \langle H\rangle$ so $t_{\rm edS}$ is typically longer than the scrambling time, unless $\epsilon_H$ rapidly departs from zero.
 Hence whereas the frequency of the mode outside the horizon has an upper bound by the finite quasi-dS lifetime $t_{\rm edS}$, it would be trans-Planckian.
  Since $(a_f/a_i)={\rm exp}[\langle H\rangle t_{\rm edS}]$, from the condition
 \dis{\Big(\frac{a_f}{a_i}\Big)\frac{1}{\omega_{\rm max}}=\frac{1}{H_f},}
 the maximal frequency $\omega_{\rm max}$ is given by  
 \dis{\omega_{\rm max}=H_f \Big(\frac{\mpl}{H_f}\Big)^{\frac{q+2}{\sqrt2 \gamma p}\frac{\langle H\rangle}{\langle \sqrt{\epsilon_H}H\rangle}}. \label{Eq:ommax}}
 This indicates that if $\langle \sqrt{\epsilon_H}H\rangle \ll \langle H\rangle$, $\omega_{\rm max}$ is much larger than $\mpl$.

While the entropic quasi-dS instability time  does not resolve the trans-Planckian problem, it is consistent with the complementarity.
This was shown in  \cite{Nomura:2011dt},  where the dS analogy of the black hole argument in  \cite{Sekino:2008he} was considered. 
More concretely, when the traveler $A$ escapes the horizon at $t_{\rm esc}$ and sends more than one bit of information after $\Delta t$, the true vacuum nucleated in dS spacetime can receive it and have a duplication of information that originally $A$ had.
This can be circumvented when the true vacuum is nucleated sufficiently late such that the comoving size of the true vacuum bubble is smaller than the comoving distance between the positions of $A$ at $t_{\rm esc}$ and  at $t_{\rm esc}+\Delta t$.
Then the time interval $\Delta T$ between the true vacuum nucleation and the escape of $A$  is bounded by  \cite{Nomura:2011dt}
\dis{\Delta T \gtrsim \frac{1}{H}\log\Big(\frac{1}{H(\Delta t)_{\rm min}}\Big).}
 If we naively apply $\Delta t \gtrsim \omega_{\rm max}^{-1}$, from \eqref{Eq:ommax}, we obtain $\Delta T \gtrsim t_{\rm edS}$.
 However, the holographic principle tells us that the amount of information that the traveler can send during $\Delta t$ is larger than $\mpl^2\Delta t^2$ and for it to contain more than one bit, $\log 2$, $\Delta t \gtrsim \mpl^{-1}$ should be satisfied.
 Hence the scrambling time $(1/H)\log(\mpl/H)$ becomes the natural bound on $\Delta T$ and since $t_{\rm edS} > \Delta T$, the true vacuum cannot have a duplication of information.
 In the case of trans-Planckian censorship conjecture, on the other hand, the complementarity issue is trivially resolved since the horizon collapses even before the quantum fluctuation becomes the mixed state   such that meaningful information is produced \cite{pricom}.
 We note that the premises of the complementarity is now under doubt \cite{Almheiri:2012rt} so the consideration above may not be relevant to the correct quantum gravity description.
  The implication of discussions in \cite{Almheiri:2012rt} to the  thermodynamic argument connecting the distance- and dS swampland conjecture, as well as the trans-Planckian issue can be the possible direction for the future studies.

  Finally, we point out that whereas the entropy contribution  $S=N^p (H/\mpl)^q$ ($q>-2$) from the descending UV degrees of freedom is an ansatz, the inequality in the form of $N^m (H/\mpl)^n <1$ ($n, m >0$) which can be found in \eqref{Eq:ineq} is quite generic.
 For example, the one loop correction to the Planck mass, or equivalently, the Newtonian constant is proportional to the number of particle species in the loop, resulting in the bound on the cutoff $\Lambda$ satisfying $N(\Lambda/\mpl)^2<{\cal O}(1)$ \cite{Veneziano:2001ah}.
 Such a bound on $\Lambda$ depending on $N$ particle species can be justified by the black hole argument as well \cite{Dvali:2007hz}, or applying the second law of thermodynamics to the Hubble entropy bound \cite{Brustein:1999ua}.
 Since $H$ is the natural cutoff scale of dS spacetime, these arguments give the inequality $N(H/\mpl)^2<{\cal O}(1)$.
 The distance conjecture states that $N$ as well as $H$ depend on the modulus $\phi$, and the entropic quasi-dS instability time obtained from this is interpreted as a time scale after which the EFT description on which the argument is based is not valid.
 
 \section{Conclusion}   
 
 The entropy argument for the dS swampland conjecture in \cite{Ooguri:2018wrx} explains the instability of quasi-dS spacetime by connecting the geometric entropy bound with the distance conjecture.
 In this work, we point out that the argument above provides the entropic quasi-dS instability time as the lifetime of quasi-dS spacetime. 
 It has a logarithmic enhancement factor $\log(\mpl/H)$, resulting in the enhancement of $\Delta \phi$ and possibly relaxation of the bound on $\mpl^2 \nabla^2 V/V$ at the top of the potential.
 The same logarithmic factor appears in  the scrambling time, which has the entropic origin as well.
 Such a connection between the entropy and characteristic time scales of (quasi-)dS spacetime implies that the features of quantum gravity  which are not reflected in  the background geometry can be explored in the thermodynamic language.
 
\vspace{5mm}
Note: After this paper was put forward in arXiv, \cite{Cai:2019dzj} reached the same conclusion for the entropic quasi-dS instability time as the lifetime of quasi-dS spacetime from the different approach. 

\vspace{5mm}
\begin{acknowledgments}

Acknowledgments:
 MS thanks Alek Bedroya, Jinn-Ouk Gong, Gary Shiu, and Dong-han Yeom for discussions and comments.

\end{acknowledgments}


\end{document}